\documentclass[
twocolumn]{aastex631}

\usepackage{amsmath,amssymb}
\usepackage[english]{babel}
\usepackage{graphicx} 
\usepackage{dcolumn} 
\usepackage{bm} 
\usepackage{makecell} 
\usepackage{threeparttable}
\usepackage{multirow} 
\usepackage{booktabs}
 
\usepackage[T1]{fontenc} 
\usepackage{color}

\begin{document}

\title{
The Nature of Gravitational Wave Events with Host Environment Escape Velocities} 

\correspondingauthor{Xi-Long Fan}
\email{xilong.fan@whu.edu.cn}

\author{Guo-Peng Li}
\affiliation{
Department of Astronomy, School of Physics and Technology, Wuhan University, Wuhan 430072, China}

\author{Xi-Long Fan}
\affiliation{
Department of Astronomy, School of Physics and Technology, Wuhan University, Wuhan 430072, China}
\date{\today}

\begin{abstract}
We propose a novel method to probe the parameters and origin channels of gravitational wave events using the escape velocities of their host environments. This method could lead to more convergent posterior distributions offering additional insights into the physical properties, formation, and evolution of the sources. 
The method provides more accurate parameter estimation for events that represent previous mergers in the hierarchical triple merger scenario and is valuable for the search for such mergers with third-generation ground-based detectors. To demonstrate this approach, we take six recently identified events in LIGO-Virgo-KAGRA data, considered as potential previous mergers in hierarchical triple mergers, as examples. The use of escape velocities results in posterior spin distributions that are concentrated near zero, aligning with the expected birth spins of first-generation black holes formed from the collapse of stars. 
The uncertainty in the posterior primary mass distribution is significantly reduced comparing with the LIGO-Virgo-KAGRA distributions, especially for events modeled under the assumption of a globular cluster origin scenario.
We rule out the possibility that GW190512, GW170729, and GW190708  originates from globular clusters as previous mergers in the hierarchical triple merger scenario.
\end{abstract}


\section{Introduction}
The LIGO-Virgo-KAGRA (LVK)~\citep{2015CQGra..32g4001L,2015CQGra..32b4001A,2012CQGra..29l4007S,2013PhRvD..88d3007A} Collaboration has significantly advanced our understanding of the Universe, opening a new window to observe cosmic phenomena that were previously inaccessible. 
Accurate estimation of the parameters of gravitational wave (GW) events is crucial for several reasons.

First, it allows us to determine the physical properties of the sources of GW waves, such as 
the masses~(e.g.,~\citealp{2022ApJ...941L..39W,2024ApJ...977...67L,2024PhRvL.133e1401L,2024MNRAS.527..298T,2025PhRvD.111b3013M,2025PhRvL.134a1401A}), 
spins~(e.g.,~\citealp{2021MNRAS.504..910T,2021PhRvD.104h3010R,2022MNRAS.514.3886M,2024A&A...692A..80P,2025arXiv250109495L}),
and orbital characteristics~(e.g.,~\citealp{2022ApJ...928L...1L,2023NatAs...7...11G,2024PhRvD.109b4024M,2024PhRvD.110f3012F,2024arXiv240101743H}) 
of binary systems. These parameters are key to understanding how compact objects like black holes (BHs) and neutron stars form and evolve. For instance, they provide crucial insights into their formation channels~\citep{2021hgwa.bookE..16M,2022PhR...955....1M}---whether these BHs result from 
isolated binary evolution~(e.g.,~\citealp{2016Natur.534..512B,2017NatCo...814906S,2019ApJ...881L...1F,2021ApJ...920...81S,2021ApJ...920L..13B,2022ApJ...928..163H,2024ApJ...965..177W,2024ApJ...974..211Z,2024arXiv241203461B}), 
dynamical interactions in galactic centers~(e.g.,~\citealp{2016MNRAS.460.3494S,2018ApJ...856..140H,2019ApJ...877...87Z,2020ApJ...894...15B,2020ApJ...901..125D,2021ApJ...917...76W,2023Univ....9..138A,2024ApJ...971L..38K,2024MNRAS.534.1634L}), 
galactic fields~(e.g.,~\citealp{2019ApJ...887L..36M,2020MNRAS.498.4924M,2022ApJ...936..184M,2022MNRAS.514.4246R,2024ApJ...965..148X,2024ApJ...972L..19S}),
globular clusters (GCs)~(e.g.,~\citealp{2019PhRvD.100d3027R,2018MNRAS.481.4775D,2020MNRAS.492.2936A,2021ApJ...921L..43Z,2024MNRAS.531..739G,2025MNRAS.538..639B,2023A&A...673A...8D}),
active galactic nuclei (AGN) disks~(e.g.,~\citealp{2021ApJ...907L..20T,2022Natur.603..237S,2022PhRvD.105f3006L,2022arXiv220406002M,2023ApJ...944L..42L,2023MNRAS.524.6015L,2023arXiv230312539G,2025arXiv250110703W,2025ApJ...979L..27L}), 
hierarchical (triple) mergers/systems~(e.g.,~\citealp{2003ApJ...598..419W,2014ApJ...784...71S,2016ARA&A..54..441N,2019ApJ...871...91Z,2020ApJ...903...67M,2021NatAs...5..749G,2019PhRvL.123r1101Y,2025ApJ...981..177L}),
and primordial black hole scenarios~(e.g.,~\citealp{PhysRevLett.116.201301,PhysRevLett.117.061101,2022EPJC...82....9W,2024JCAP...08..030H}).

Second, accurate parameter estimation (PE) is essential for testing general relativity in the strong-field regime~\citep{2021Univ....7..497K}. This provides an opportunity to search for potential deviations from the theory, offering new insights into the limits of our current understanding of gravity.  
Finally, precise PE enhances our ability to localize GW sources in the sky. This is particularly important for multi-messenger astronomy, where GW detections are followed up by electromagnetic observations across the electromagnetic spectrum, spanning from radio waves to gamma rays~\citep{2020PhRvL.124y1102G,2023NatAs...7..579M,2024arXiv240910651K,2025A&A...693A..84K,2025arXiv250313884G}. 
Identifying the host galaxies of these events provides additional context, such as the environment in which the mergers occurred, and sheds light on the role of various astrophysical processes in shaping these events. Moreover, accurate localization is vital for identifying potential counterparts, such as electromagnetic counterparts to binary BH (BBH) mergers in AGN disks~\citep{2020PhRvL.124y1102G,2023ApJ...942...99G,2022MNRAS.514.2092V,2024arXiv240505318V}.  
In summary, accurate PE is a cornerstone of GW astronomy, enabling a deeper understanding of the Universe and driving advancements in both theoretical and observational astrophysics.

PE in GW astronomy faces several significant challenges. One primary difficulty arises from the inherent noise in data collected by GW detectors~\citep{2020CQGra..37e5002A}. These detectors are extremely sensitive, capable of measuring spacetime distortions on the order of one-thousandth the diameter of a proton. However, this sensitivity makes them highly susceptible to various noise sources, including seismic activity, thermal fluctuations, and instrumental artifacts. Distinguishing genuine GW signals from background noise requires sophisticated, computationally intensive data analysis techniques.  
Another major challenge is parameter degeneracy in GW signals. For example, similar waveform features can arise from different combinations of masses and spins of merging objects, complicating efforts to disentangle their individual contributions. This degeneracy often leads to large uncertainties in estimated parameters, limiting our ability to precisely characterize the sources. Additionally, the limited bandwidth of current detectors restricts the observable portion of the GW signal, further complicating the PE process.

To address these challenges, additional constraints and complementary observations are crucial. One promising approach involves incorporating information from the host environments of GW sources~\citep{2014ApJ...795...43F,2017PhRvL.119r1102F,2024arXiv241014663M}. For instance, the escape velocity of a host environment provides a natural constraint on the recoil (or ``kick'') velocity imparted to the remnant BH due to the linear momentum carried away by GW radiation~\citep{2007PhRvL..98i1101G}. Previous studies have made significant progress in this area by calculating the kick velocities of remnant BHs for GW events~\citep{2021ApJ...918L..31M,2022PhRvL.128s1102V}. For example, GW190814~\citep{2020ApJ...896L..44A} was found to have a relatively small kick velocity of $\sim$$74^{+10}_{-7}\,{\rm km\,s^{-1}}$~\citep{2021ApJ...918L..31M}, whereas GW200129\_065458~\citep{2023PhRvX..13d1039A} exhibited a much larger kick velocity of $\sim$$1542^{+747}_{-1098}\,{\rm km\,s^{-1}}$~\citep{2022PhRvL.128s1102V} (both at 90\% credibility). These studies have highlighted the implications of kick velocities for understanding GW events, particularly in linking remnant BH properties to their astrophysical contexts. However, most of this prior work primarily focused on quantifying kick velocities and their implications without delving deeper into the nature of GW events or systematically integrating kick constraints with other astrophysical and observational information.

In this study, we propose a novel method to constrain the parameters of GW events detected by LVK using the escape velocities of host environments. This approach not only provides a more detailed understanding of the recoil dynamics but also offers new insights into the formation channels, environments, and physical processes governing GW sources. By incorporating escape velocity as a constraint, we can more precisely define the parameter space of GW events, significantly improving the accuracy of parameter estimation. Furthermore, we find that escape velocity serves as a constraint on the origin channels of GW events by excluding channels with distributions that deviate unreasonably from those reported by the LVK Collaboration. This methodology bridges the gap between GW observations and astrophysical theory, enabling a deeper exploration of the nature of GW events and their broader cosmological implications. Future research can build upon this framework by integrating additional astrophysical information to further enhance the accuracy and reliability of PE. This approach holds promise for unraveling the complexities of GW sources and advancing our understanding of the Universe.

The rest of this paper is structured as follows. 
Section~\ref{sec:HTM} introduces hierarchical triple mergers, followed by a detailed description of our methodology in Section~\ref{sec:RE}.  
Our results are presented in Section~\ref{sec:KV}, which includes the distribution of kick velocities and the retention fraction of GW samples, and in Section~\ref{sec:PD}, which covers the distributions of GW parameters. 
Sections~\ref{sec:GW170818} and~\ref{sec:GW190512} examine two events (GW170818 and GW190512) as illustrative examples.  
The implications for formation channels are discussed in Section~\ref{sec:IOC}.  
Section~\ref{sec:discussion} addresses the limitations of our approach.
Finally, we summarize our conclusions in Section~\ref{sec:conclusions}.

\section{Hierarchical triple mergers}\label{sec:HTM}

Our method is particularly well-suited for GW events that are potentially the result of previous mergers in hierarchical triple merger scenarios. 
The hierarchical triple merger is a specific BBH merger scenario~\citep{2018MNRAS.476.1548S,2019MNRAS.482...30S}, proposed to occur predominantly in dynamical formation channels such as star clusters and AGN disks since these channels are efficient at producing hierarchical mergers~(e.g.,~\citealp{2019PhRvL.123r1101Y,2020MNRAS.494.1203M,2021Symm...13.1678M,2021MNRAS.505..339M,2023PhRvD.107f3007L,2023Univ....9..138A}). 
In this scenario, three BHs form a gravitationally bound three-body system, where mutual interactions facilitate the inspiral of two BHs~\citep{PhysRevD.77.101501,PhysRevD.77.024034}. Initially, two of the BHs merge to emit a GW signal, leaving behind a remnant BH. This remnant BH subsequently forms a new binary with the third BH and merges to produce a second GW signal. Under certain orbital configurations, particularly in AGN disks due to the frequent occurrence of hierarchical mergers and the role of gas in facilitating these events~\citep{2019PhRvL.123r1101Y,2020MNRAS.494.1203M,2022PhRvD.105f3006L}, both mergers could be observed within a timescale of a few years~\citep{2017ApJ...835..165B,2019ApJ...878...85S}.

Hierarchical triple systems are widely used to explain certain special GW events featuring significantly asymmetric component masses, components in the lower/upper ``mass gap'', or high spins, which are promising candidates for hierarchical mergers.  
For example, GW190412~\citep{2020PhRvD.102d3015A}, GW190521~\citep{2020PhRvL.125j1102A}, and GW190814~\citep{2020ApJ...896L..44A} have been investigated in the context of hierarchical triple systems, suggesting that such events could be produced via hierarchical mergers in multiples~\citep{2021MNRAS.502.2049L,2021MNRAS.500.1817L,2023arXiv230709097G}.  
In particular, \citet{2023arXiv230709097G} investigated whether GW observations in the lower mass gap favor a hierarchical triple origin and showed that a neutron star merger origin for the lighter components in GW190814~\citep{2020ApJ...896L..44A} and GW200210\_092254~\citep{2023PhRvX..13d1039A} is favored over various mass distributions in the lower mass gap.  
Further, \citet{2024ApJ...965...80G} discussed the prospects for identifying hierarchical triple mergers with third-generation ground-based detectors~\citep{2010CQGra..27s4002P,2010CQGra..27h4007P,2017CQGra..34d4001A}, and \citet{2023MNRAS.523.4113T} proposed testing Hawking's area theorem~\citep{PhysRevLett.26.1344} using inspiral signals from hierarchical triple mergers.

Recent searches~\citep{2020MNRAS.498L..46V,2021ApJ...907L..48V} for hierarchical triple mergers have identified several significant merger pairs based on a frequentist $p$-value assignment using a test statistic, yielding intriguing candidate families and insights into their astrophysical implications.  
Notably, six GW events have been identified as potential previous mergers in this scenario~\citep{2021ApJ...907L..48V}: GW170729, GW170818, GW170823, GW190512, GW190514, and GW190708, within the first and second gravitational-wave transient catalogs (GWTC;~\citealp{2019PhRvX...9c1040A,2021PhRvX..11b1053A}) (see their significances in Table~1 of~\citealp{2021ApJ...907L..48V}).  
In particular, they found that GW190514 and GW170729 are the two most plausible predecessors to GW190521 (with significances $p \lesssim 0.14$ and $\lesssim 0.32$, respectively), under the assumption that the upper bound of the mass distribution of first-generation BHs is $50\,{M_\odot}$.  
The GW190519-GW170818 pair becomes the most significant (with $\sim$$3\sigma$ significance) when the upper bound is increased to 60, 70, or $100\,{M_\odot}$.

In addition, their search revealed an interesting pair~\citep{2021ApJ...907L..48V}: GW190521-GW190514 as a potential hierarchical triple merger in an AGN.  
This is supported by the fact that AGN J124942.3+344929, from which a candidate electromagnetic counterpart ZTF19abanrhr was detected by ZTF after GW190521~\citep{2020PhRvL.124y1102G}, lies within the 90\% sky localization of GW190514 and may also be associated with it~\citep{2023ApJ...942...99G}.  
\citet{li2025multimessengerhierarchicaltriplemerger} recently investigated this plausible hierarchical triple merger scenario, positing GW190514 as the precursor merger to GW190521, with one or both potentially associated with ZTF19abanrhr, providing evidence for the first hierarchical triple merger with an EM counterpart in the AGN formation channel.

\section{Retention vs. escape}\label{sec:RE}
A critical condition for hierarchical triple mergers is the retention of remnant BHs in their host environments. This requires that the kick velocities imparted to the remnants during the previous merger must be less than the escape velocities of their host environments~\citep{2021ApJ...918L..31M,2022A&A...666A.194L}. The escape velocity is defined as the minimum speed required for an object to escape the gravitational influence of its host celestial body, and it varies significantly across different astrophysical environments. For these six GW events, the posterior kick velocity samples from PE can be compared to the host environment's escape velocity. Samples exceeding the escape velocity can be excluded, as this scenario necessitates that the remnant BHs remain gravitationally bound to their host environments to participate in subsequent mergers (assuming that dynamical N-body interactions do not play a dominant role in ejecting heavy remnant BHs). This reads following
Bayes' theorem: 
\begin{equation}
    p({\boldsymbol{\lambda},V_{\rm kick}|d}) 
    \propto 
    \mathcal{L}(d|\boldsymbol{\lambda},V_{\rm kick}) \, 
    \pi(\boldsymbol{\lambda}) \,
    \pi(V_{\rm kick}|\boldsymbol{\lambda}) \,
    \label{eq:1}
\end{equation}
where $p({\boldsymbol{\lambda},V_{\rm kick}|d})$ is the posterior probability distribution of the 15-dimensional binary parameters $\boldsymbol{\lambda}$ for quasicircular BBHs and the remnant BH kick velocity $V_{\rm kick}$ given the observed data $d$, $\mathcal{L}(d|\boldsymbol{\lambda},V_{\rm kick})$ is the likelihood function of the data given $\boldsymbol{\lambda}$ and $V_{\rm kick}$ setting to be $\mathcal{L}(d|\boldsymbol{\lambda})$. Note that, in the future one could assign more sophisticated models taking into account the $V_{\rm kick}$ in the GW data analysis processes. $\pi(\boldsymbol{\lambda})$ is the prior adopted in current GW data analysis. $\pi(V_{\rm kick}|\boldsymbol{\lambda})$ is the conditional prior probability distributions for $V_{\rm kick}$ given by:
\begin{equation}
\pi(V_{\rm kick}|\boldsymbol{\lambda}) = \left\{
\begin{aligned}
&\frac{1}{V^{\rm max}_{\rm esc}}, &{\rm if~} V_{\rm kick}(\boldsymbol{\lambda}) < V^{\rm max}_{\rm esc}\,, \\
&0, &{\rm others}\,,
\end{aligned}
\right.
\label{eq:2}
\end{equation}
where $V^{\rm max}_{\rm esc}$ is the upper bound for the host environment escape velocity range, and $V_{\rm kick}(\boldsymbol{\lambda})$ is the kick velocity, which depends only on the intrinsic parameters $\{m1,m2,\boldsymbol{\chi}_1,\boldsymbol{\chi}_2\}$ in $\boldsymbol{\lambda}$, where index 1 (2) corresponds to the primary (second) BH, with $m_{1,2}$ the masses and $\boldsymbol{\chi}_{1,2}$ the dimensionless spins. We employ the numerical relativity fitting formulas~\citep{2007ApJ...659L...5C} to infer the kick $V_{\rm kick}(\boldsymbol{\lambda})$, which the $\boldsymbol{\Lambda}$ posterior samples (we directly use the posterior samples from GWTC-2.1,~\citealp{2024PhRvD.109b2001A}) yield  the posterior distribution of $V_{\rm kick}$.

\section{Kick velocities}\label{sec:KV}
We compute the posterior distributions of the kick velocities for the remnants of six GW events and deduce the fractions of posterior samples retained within their host environments. The retention fraction can also be interpreted as the probability that the remnant BH of a merger remains bound to its host environment. This retention probability is critical for determining whether these GW events can facilitate hierarchical mergers, as only mergers with at least one retained remnant BH are considered part of hierarchical sequences.

\begin{table}
\caption{
The kick velocities and the upper limits of retention fractions (or probabilities) for the posterior samples of the six GW events identified as previous mergers in the hierarchical triple merger scenario.  
}\resizebox{\linewidth}{!}{
\begin{tabular}{lcccc}
\toprule
\multirow{2}{*}{Event} & \multirow{2}{*}{$V_{\rm kick}\,[{\rm km\,s^{-1}}]$} & \multicolumn{3}{c}{Fraction} \\
&&AGN disks & NSCs & GCs \\
\midrule
GW170729 & $915.38^{+1223.40}_{-693.32}$ & $0.5487$ & $0.3164$ & $0.0009$ \\
GW170818 & $211.92^{+146.81}_{-130.77}$ & $1.0000$ & $1.0000$ & $0.0882$ \\
GW170823 & $1029.04^{+1229.74}_{-776.23}$ & $0.4814$ & $0.2274$ & $0.0033$ \\
GW190512 & $526.79^{+843.99}_{-350.18}$ & $0.8526$ & $0.5771$ & $0.0046$ \\
GW190514 & $819.23^{+1115.82}_{-596.36}$ & $0.6290$ & $0.3263$ & $0.0042$ \\
GW190708 & $194.23^{+179.29}_{-102.92}$ & $1.0000$ & $0.9983$ & $0.0630$ \\
\bottomrule
\end{tabular}}
\label{tab1}
\begin{tablenotes} 
\item 
{\bf Note.}
Column 1: Names of the GW events.  
Column 2: Median and 90\% credible intervals for the kick velocities.  
Columns 3-5: Upper limits of the retention fractions for the posterior samples, assuming the mergers occur in AGN disks, NSCs, or GCs.  
\end{tablenotes}
\end{table}
Table~\ref{tab1} summarizes the kick velocities inferred for the six GW events, which span a wide range: $\sim$$50{\text-}2500\,{\rm km\,s^{-1}}$. 
For comparison, escape velocities in dense astrophysical environments hosting BHs range from $\mathcal{O}(10)\,{\rm km\,s^{-1}}$ for GCs~\citep{2016ApJ...831..187A} to $\mathcal{O}(100)\,{\rm km\,s^{-1}}$ for nuclear star clusters (NSCs)~\citep{2016ApJ...831..187A}, and up to $\sim$$1000\,{\rm km\,s^{-1}}$ in AGN disks. 
We therefore assume that the upper bounds of these escape velocity ranges in GCs, NSCs, and AGN disks are $\sim$$100\,{\rm km\,s^{-1}}$, $\sim$$600\,{\rm km\,s^{-1}}$, and $\sim$$1000\,{\rm km\,s^{-1}}$, respectively~\citep{2022PhRvL.128s1102V}.
Using the upper bounds, we calculate the upper limits for the retention fractions in AGN disks, NSCs, and GCs. For instance, in GCs, samples with kick velocities exceeding $\sim$$100\,{\rm km\,s^{-1}}$ are excluded. We find that the retention fraction decreases as the escape velocity decreases, indicating that mergers in AGN disks require the fewest samples to be removed, while mergers in GCs require the most.

Among the six events, GW170818 and GW190708 exhibit relatively low kick velocities ($\sim$$200\,{\rm km\,s^{-1}}$), making them the most suitable candidates for hierarchical triple mergers. This aligns with the findings of \citet{2021ApJ...907L..48V}, which identified the pairs GW190519-GW170818 and GW190915-GW190708 as likely hierarchical triple mergers. GW170818 and GW190708 are retained with probabilities up to $\sim$100\% in AGN disks and NSCs. However, in GCs, their retention probabilities drop significantly to $\sim$8.8\% and $\sim$6.3\%, respectively. This suggests that GCs, due to their low escape velocities, are inefficient at producing hierarchical mergers and are unlikely sites for the observed hierarchical triple mergers.

The remaining four events have relatively high kick velocities, ranging from $\sim$$527\,{\rm km\,s^{-1}}$ to $\sim$$1029\,{\rm km\,s^{-1}}$. Assuming the escape velocity of the host environment is at the upper bound for GCs, their retention probabilities are $\lesssim$0.04\%. In NSCs, the retention probabilities improve to $\leq$50\% (but $\sim$57.7\% for GW190512 with $V_{\rm kick} \sim 527\,{\rm km\,s^{-1}}$). In AGN disks, their retention probabilities exceed $\sim$50\% (but $\sim$48.1\% for GW170823 with $V_{\rm kick} \sim 1029\,{\rm km\,s^{-1}}$). These findings indicate that AGN disks, followed by NSCs, are the most promising sites for hierarchical mergers and the observation of hierarchical triple mergers, with respect with GCs.

\begin{figure}
\centering
\includegraphics[width=8cm]{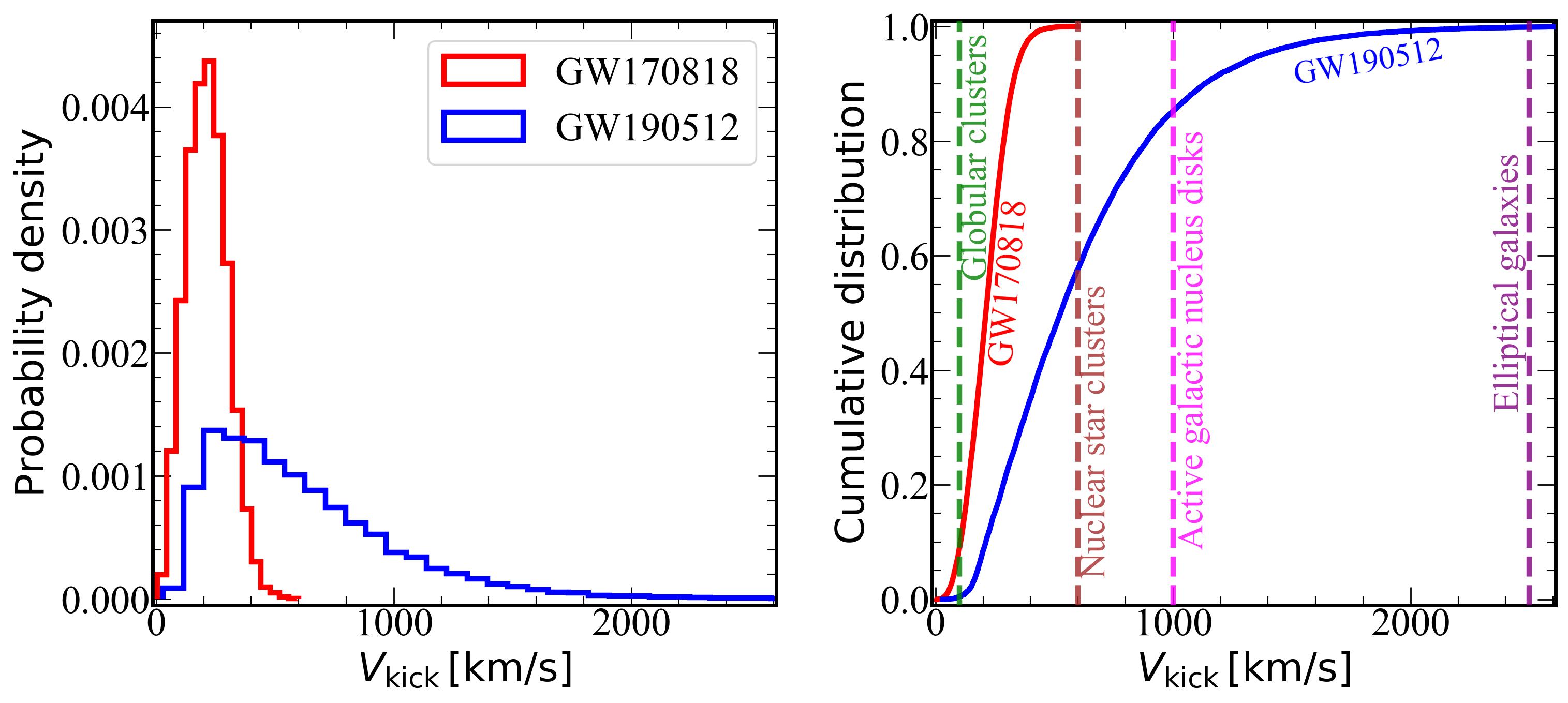}
\caption{
The probability distributions (left) and cumulative distributions (right) of the kick velocities for GW170818 (red) and GW190512 (blue). The vertical dotted lines represent the upper bounds of escape velocity ranges for various types of host environments, provided for comparison. The upper limit of the retention fraction (or probability) for the merger remnant is determined by the intersection of these lines with the posterior cumulative distributions.  
}
\label{kick} 
\end{figure}
Figure\,\ref{kick} illustrates the probability and cumulative distributions of kick velocities for GW170818 and GW190512, alongside fiducial escape velocities for GCs~\citep{2016ApJ...831..187A}, NSCs~\citep{2016ApJ...831..187A}, AGN disks, and giant elliptical galaxies~\citep{2004ApJ...607L...9M}. GW170818 exhibits lower kick velocities ($211.92^{+146.81}_{-130.77}\,{\rm km\,s^{-1}}$, 90\% credible interval), making it more likely to be retained by its host environment compared to GW190512, which the latter has higher kick velocities ($526.79^{+843.99}_{-350.18}\,{\rm km\,s^{-1}}$). Consequently, GW170818 has higher retention probabilities, reaching 100\% in AGN disks and NSCs, whereas GW190512 retains probabilities of $\sim$85.3\% and $\sim$57.7\%, respectively, in the same environments. This suggests GW170818 is a stronger candidate for the part of hierarchical triple mergers, consistent with its significance~\citep{2021ApJ...907L..48V}.

\section{Posterior distributions}\label{sec:PD}
\begin{table*}
\caption{
The posterior distributions of GW170818 and GW190512 for various types of host environments, including AGN disks, NSCs, and GCs. 
}\resizebox{\linewidth}{!}{
\begin{tabular}{ccccccccccc}
\toprule
Event & Host & $m_1\,[M_{\odot}]$ & $m_2\,[M_{\odot}]$ & $\chi_1$ & $\chi_2$ & $q$ & $\mathcal{M}\,[M_{\odot}]$ & $\chi_{\rm eff}$ & $\chi_{\rm p}$ \\
\midrule
\multirow{4}{*}{GW170818} & $\cdots$ & $34.78^{+6.52}_{-4.19}$ & $27.60^{+4.15}_{-5.10}$ & $0.52^{+0.41}_{-0.47}$ & $0.48^{+0.46}_{-0.42}$ & $0.80^{+0.18}_{-0.24}$ & $26.83^{+2.32}_{-2.00}$ & $-0.06^{+0.19}_{-0.22}$ & $0.56^{+0.34}_{-0.41}$ \\
 & AGN disks &  $34.78^{+6.52}_{-4.19}$ & $27.60^{+4.15}_{-5.10}$ & $0.52^{+0.41}_{-0.47}$ & $0.48^{+0.46}_{-0.42}$ & $0.80^{+0.18}_{-0.24}$ & $26.83^{+2.32}_{-2.00}$ & $-0.06^{+0.19}_{-0.22}$ & $0.56^{+0.34}_{-0.41}$ \\
 & NSCs  & $34.78^{+6.52}_{-4.19}$ & $27.60^{+4.15}_{-5.10}$ & $0.52^{+0.41}_{-0.47}$ & $0.48^{+0.46}_{-0.42}$ & $0.80^{+0.18}_{-0.24}$ & $26.83^{+2.32}_{-2.00}$ & $-0.06^{+0.19}_{-0.22}$ & $0.56^{+0.34}_{-0.41}$ \\
 & GCs &  $33.31^{+4.14}_{-3.36}$ & $28.59^{+3.47}_{-3.93}$ & $0.17^{+0.35}_{-0.16}$ & $0.24^{+0.43}_{-0.21}$ & $0.87^{+0.12}_{-0.18}$ & $26.72^{+2.44}_{-1.87}$ & $-0.04^{+0.14}_{-0.21}$ & $0.19^{+0.35}_{-0.14}$ \\
\multirow{4}{*}{GW190512}  & $\cdots$ & $23.15^{+5.64}_{-5.61}$ & $12.53^{+3.49}_{-2.55}$ & $0.20^{+0.49}_{-0.18}$ & $0.40^{+0.51}_{-0.37}$ & $0.54^{+0.36}_{-0.18}$ & $14.56^{+1.36}_{-0.94}$ & $0.02^{+0.13}_{-0.14}$ & $0.26^{+0.41}_{-0.20}$ \\
 & AGN disks & $23.50^{+5.43}_{-5.66}$ & $12.38^{+3.45}_{-2.47}$ & $0.17^{+0.34}_{-0.15}$ & $0.37^{+0.51}_{-0.33}$ & $0.53^{+0.35}_{-0.17}$ & $14.57^{+1.34}_{-0.95}$ & $0.03^{+0.12}_{-0.13}$ & $0.22^{+0.27}_{-0.17}$ \\
 & NSCs  & $23.84^{+5.24}_{-5.68}$ & $12.23^{+3.51}_{-2.40}$ & $0.14^{+0.22}_{-0.12}$ & $0.29^{+0.49}_{-0.26}$ & $0.51^{+0.34}_{-0.16}$ & $14.59^{+1.33}_{-0.97}$ & $0.03^{+0.12}_{-0.12}$ & $0.17^{+0.17}_{-0.13}$ \\
 & GCs & $18.84^{+2.52}_{-2.01}$ & $15.68^{+2.26}_{-1.61}$ & $0.03^{+0.10}_{-0.03}$ & $0.04^{+0.08}_{-0.03}$ & $0.84^{+0.13}_{-0.12}$ & $15.08^{+1.09}_{-1.09}$ & $-0.00^{+0.07}_{-0.05}$ & $0.03^{+0.08}_{-0.02}$ \\
\bottomrule
\end{tabular}}
\label{tab2}
\begin{tablenotes} 
\item 
{\bf Note.}
Column 1: Name of the GW events.  
Column 2: Host environments (AGN disks, NSCs, or GCs).  
Columns 3-10: Median and 90\% credible intervals for the posterior distributions of $m_1$, $m_2$, $\chi_1$, $\chi_2$, $q$, $\mathcal{M}$, $\chi_{\rm eff}$, and $\chi_{\rm p}$. These values are obtained based on the upper limits for the retention fractions of the posterior samples retained under the assumption that the merger occurred in specific host environments.  
`$\cdots$' denotes that the posterior distributions are sourced from GWTC-2.1~\citep{2024PhRvD.109b2001A}.  
\end{tablenotes}
\end{table*}
Table~\ref{tab2} takes GW170818 and GW190512 as examples and lists the posterior distributions ($m_1, m_2, \chi_1, \chi_2, q, \mathcal{M}, \chi_{\rm eff},~{\rm and}~\chi_{\rm p}$), assuming these events occur within AGN disks, NSCs, or GCs. The distributions reflect the constraints imposed by the upper bounds of the escape velocity ranges of these environments (used in Table~\ref{tab1}), illustrating the impact of host environment characteristics on the inferred parameters. We observe that as the retention fraction decreases, the distributions of the masses and spins shift toward smaller values. This trend arises because symmetric masses and lower spins result in smaller kick velocities, which are more likely to be retained in environments with lower escape velocities.

Restricting the posterior samples based on the escape velocities of host environments leads to more convergent distributions. These constrained distributions carry additional information that aids in further determining the physical properties, formation, and evolution of the GW sources. However, as more posterior samples are excluded, the resulting distributions may deviate further from the posterior distribution obtained by the LVK Collaboration, which does not apply such priors. 
Significant deviations between the LVK distribution and the escape-velocity-constrained distribution can help exclude certain origin channels for GW events, assuming they are previous mergers in the hierarchical triple merger scenario and remain bound to their hosts.
To illustrate this, we analyze GW170818 and GW190512 in detail, focusing on their constrained distributions and potential origin channels.

\section{GW170818: posterior distributions and host environments}\label{sec:GW170818}
\begin{figure}
\centering
\includegraphics[width=8cm]{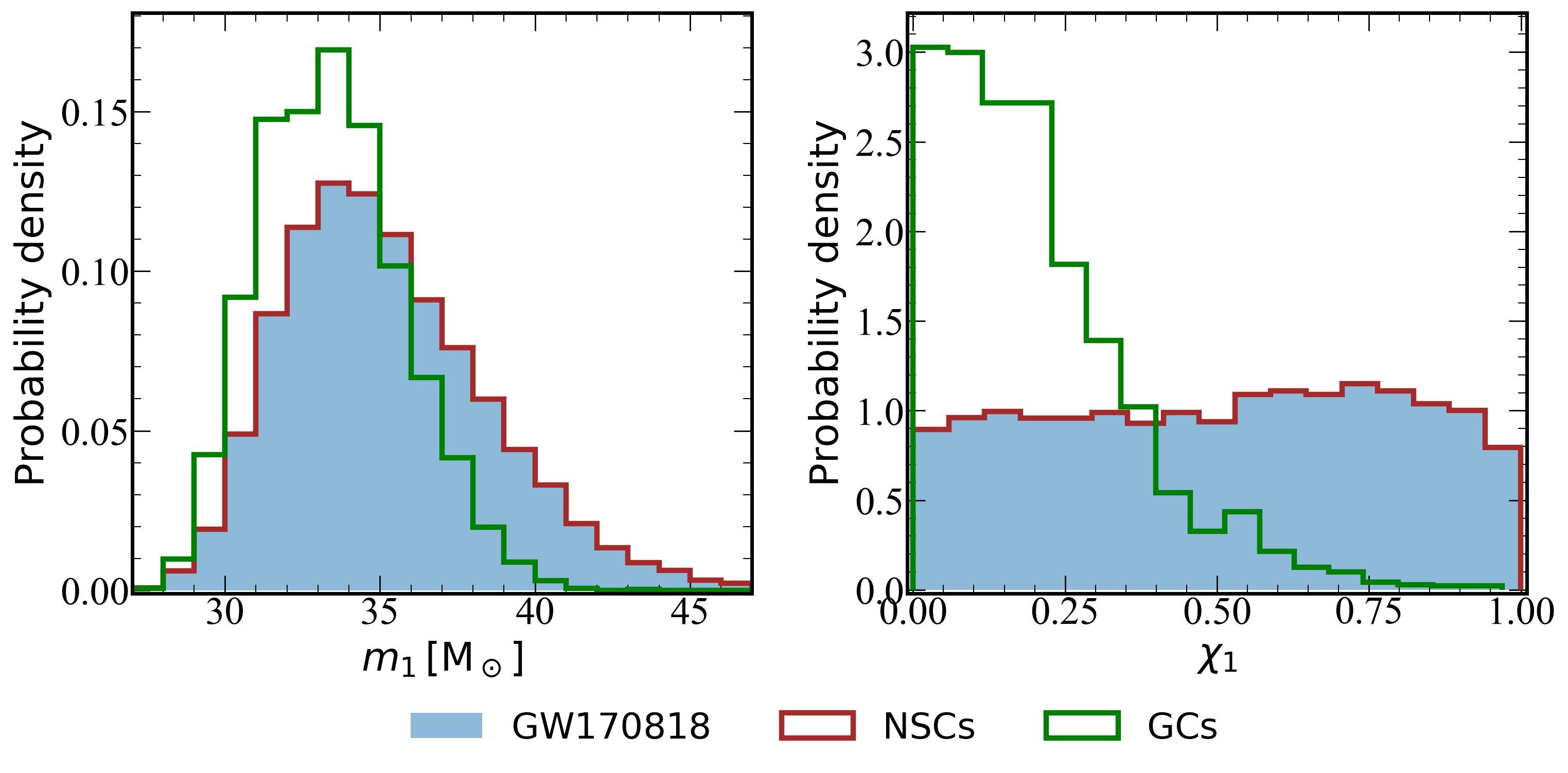}
\caption{
The posterior probability distributions for $m_1$ and $\chi_1$ of GW170818. The empty histograms represent the distributions assuming the merger takes place inside NSCs (green) or GCs (red), constrained by the upper bounds of the escape velocity ranges $\sim$$600\,{\rm km\,s^{-1}}$ and $\sim$$100\,{\rm km\,s^{-1}}$, respectively. For comparison, the filled histograms (blue) illustrate the distributions from GWTC-2.1.  
}
\label{GW170818} 
\end{figure}
Figure\,\ref{GW170818} compares the posterior probability densities of the primary mass ($m_1$) and spin magnitude ($\chi_1$) for GW170818 from the GWTC-2.1 catalog~\citep{2024PhRvD.109b2001A} and those constrained by the escape velocities of NSCs and GCs. We assume that GW170818 originates from NSCs (the upper bounds of the escape velocity ranges $\sim$$600\,{\rm km\,s^{-1}}$) or GCs (that $\sim$$100\,{\rm km\,s^{-1}}$). We find that if GW170818 originates from GCs, $\sim$91\% of the posterior samples must be excluded due to its relatively high retention fraction in GCs. Consequently, the primary mass is better constrained from $34.78^{+6.52}_{-4.19}\,M_{\odot}$ to $33.31^{+4.14}_{-3.36}\,M_{\odot}$, improving the uncertainty by $\sim$30.1\% to $\sim$22.5\%. The spin distribution becomes more concentrated around zero ($\chi_1=0.17^{+0.35}_{-0.16}$), consistent with the expectation for BHs formed from stellar collapse. In contrast, the GWTC-2.1 spin distribution is nearly uniform and lacks significant information. When constrained using the upper escape velocity for NSCs, the distribution overlaps entirely with the GWTC-2.1 distribution. This is expected, as GW170818's kick velocity ($211.92^{+146.81}_{-130.77}\,{\rm km\,s^{-1}}$) is well below the NSC escape velocity upper bound. These results suggest that GW170818 is likely retained in NSCs or AGN disks and may originate from such environments, while its retention in GCs is highly improbable. Meanwhile, if GW170818 originates from GCs, more convergent distributions can be obtained.

\section{GW190512: posterior distributions and host environments}\label{sec:GW190512}
\begin{figure}
\centering
\includegraphics[width=8cm]{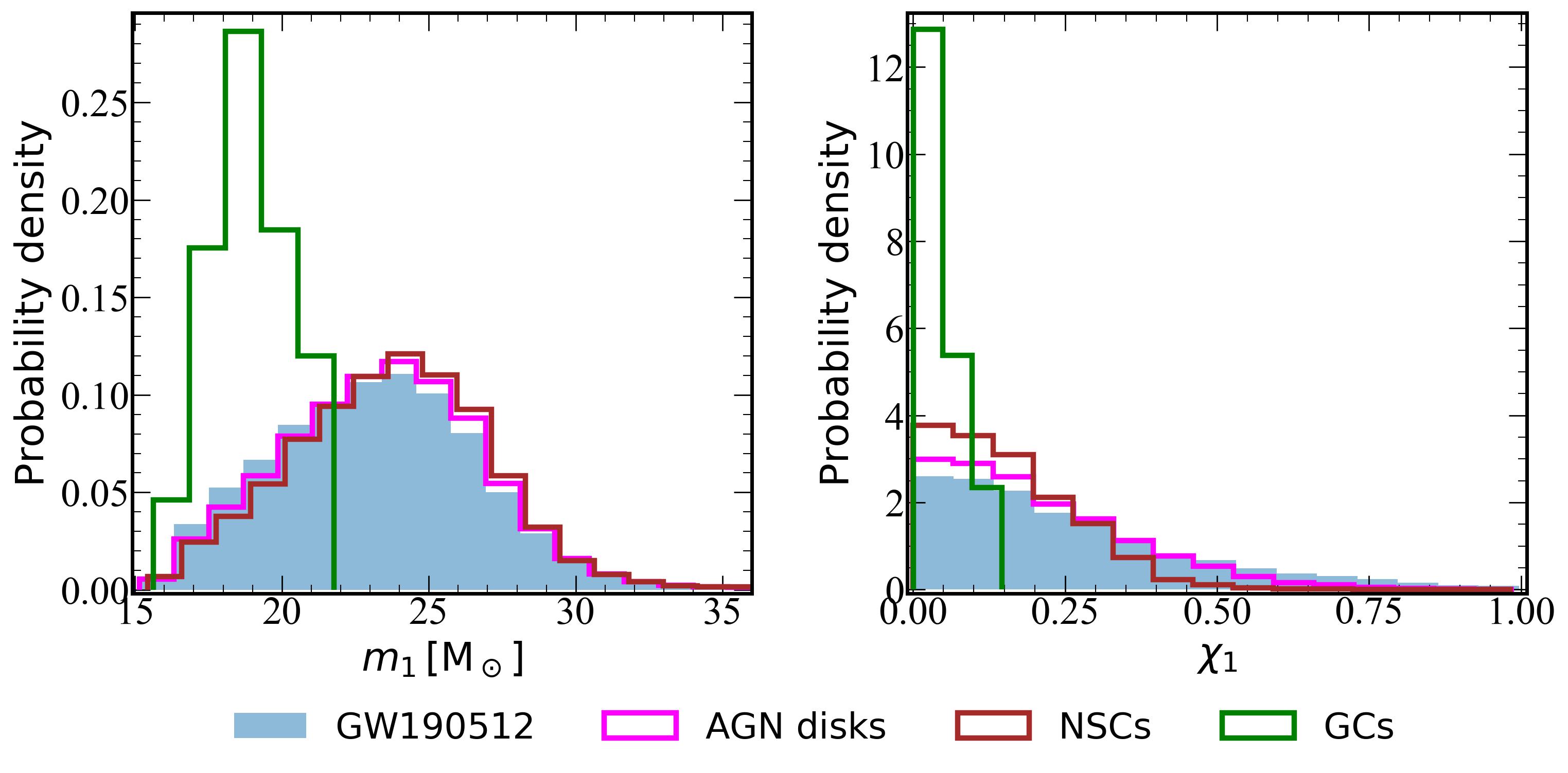}
\caption{
Same as Figure~\ref{GW170818}, but for GW190512. The empty histogram (magenta) represents the posterior distribution assuming the merger takes place inside AGN disks, constrained by the upper bound of the escape velocity range. This is shown due to the higher kick velocity of GW190512, $526.79^{+843.99}_{-350.18}\,{\rm km\,s^{-1}}$, corresponding to the upper limit of the retention fraction in AGN disks being $\sim$85.3\%.
}
\label{GW190512} 
\end{figure}
Figure\,\ref{GW190512} illustrates the posterior probability densities of the primary mass and spin magnitude for GW190512. Assuming the event originates from AGN disks or NSCs, we observe: (1) GW190512's higher kick velocity ($526.79^{+843.99}_{-350.18}\,{\rm km\,s^{-1}}$) results in retention fractions of $\sim$85.3\% and $\sim$57.7\% in AGN disks and NSCs, respectively. As a result, the constrained distributions are only slightly narrower than those in the GWTC-2.1 catalog; (2) The GW190512 spin distribution from GWTC-2.1 is already well-constrained and concentrated around zero, unlike GW170818. This leaves less room for further improvement through escape velocity constraints. If GW190512 originates from a GC, a noticeable deviation emerges between the primary mass distributions obtained from GWTC-2.1 and those constrained by GC escape velocities. 
This deviation (due to the larger $V_{\rm kick}$) suggests that GW190512 is unlikely to originate from a GC, assuming it is a previous merger in the hierarchical triple merger scenario and remains bound to its host.
We note, however, that this does not necessarily imply the merger could not occur in a GC.

\section{Implications for origin channels}\label{sec:IOC}
These findings indicate that GW170818 and GW190512 likely originate from environments with higher escape velocities, such as NSCs or AGN disks, while GCs are less favorable. Specifically, the significant deviation in the constrained distributions of GW190512 rules out GCs as its origin channel. This result further supports the use of escape velocity constraints to probe the formation and evolutionary history of GW sources, particularly in hierarchical triple merger scenarios. Conversely, once the merger comes from GCs, we can obtain its more effective posterior information.

\section{Discussion}\label{sec:discussion}
%

Our analysis assumes the condition $V_{\rm kick} < V_{\rm esc}$ to constrain previous mergers in the hierarchical triple merger scenario to remain bound to their host environments (see Section~\ref{sec:RE}).  
For example, when constraining BBH merger parameters, we restrict their posterior samples based on the escape velocities of host environments. This is based on two assumptions: (1) all mergers with $V_{\rm kick} > V_{\rm esc}$ are ejected from the host, and (2) once ejected, the remnant will no longer be bound to a tertiary companion and thus cannot undergo a subsequent hierarchical merger.  
However, these assumptions are not necessarily valid. First, a merger remnant may remain bound to a close tertiary even if $V_{\rm kick}$ is sufficient to eject the system from its host. Conversely, retention in the host environment does not guarantee that a tertiary companion remains bound post-kick, eventually merging with the remnant.  
For instance, in the GC channel, a significant fraction ($\sim$50\%) of BBH mergers can undergo multiple hardening encounters before being ejected from the cluster and later merge in the galactic field~\citep{2010MNRAS.407.1946D,2016PhRvD..93h4029R,2019PhRvD..99f3003K}. 
These ejected mergers are excluded from our hierarchical triple merger sample, requiring further analysis to distinguish their complex dynamics and identify events compatible with our framework.  
Moreover, we implicitly assume instantaneous ejection without additional interactions during escape (e.g., three-body encounters or gas drag in AGN disks), which may affect tertiary binding. The host's relaxation timescale ($>$$10^9\,{\rm yr}$;~\citealp{2012ApJ...759...52D}) also determines whether ejected systems can still merge before disruption. Even ejected triples may merge within this timescale.  
Future work should incorporate full dynamical histories of triples, including post-ejection evolution. In this study, however, we provide a novel framework to demonstrate how environment-informed priors can refine GW parameter estimation, despite incomplete coverage of hierarchical triple merger scenarios.

Our analysis adopts the upper bounds of $V_{\rm esc}$ ranges for different environments to constrain GW parameters, regardless of the fact that actual escape velocities depend on local properties.
We emphasize that $V_{\rm esc}$ varies spatially within each environment. For example, in a GC, $V_{\rm esc}$ can differ significantly between the core and the halo. Such gradients imply that retention probabilities depend on the merger location: remnants formed in the core may remain bound even with kicks exceeding the halo's $V_{\rm esc}$, while those in the outskirts may escape despite sub-threshold kicks.
As a comparison, we recalculated retention fractions and constrained parameters for various $V_{\rm kick}$ values across host environments (see Table~\ref{tab3} in Appendix~\ref{appB}).

Our parameter constraints and origin inference are performed under two different assumptions: with and without prior knowledge of the host environment.  
We first assume that the events are previous mergers in the hierarchical triple merger scenario without prior knowledge of their host environment. Kick velocities and retention fractions are used to statistically infer plausible formation environments (see Figure~\ref{kick} and Table~\ref{tab1}), although this does not uniquely identify the host environment, as multiple dynamical channels may share comparable escape velocities~\citep{2016ApJ...831..187A}.  
We then assume that mergers occur within a specific environment, imposing an upper bound on kick velocities based on the escape velocity of that environment. This allows us to compute PE under a physically motivated prior (see Figures~\ref{GW170818} and~\ref{GW190512}, and Table~\ref{tab2}).  
However, this improvement is contingent on the validity of the assumed formation environment (e.g., GCs).  
Developing a self-consistent framework combining environment-agnostic retention probabilities with environment-specific parameter estimation in the future could help resolve formation channels while preserving measurement accuracy.

We note that the improvement in parameter measurement accuracy relies on the assumptions of a specified host environment and a confidently identified hierarchical triple merger. However, current GW detectors lack the precision to directly pinpoint the host environments of BBH mergers. 
In particular, the host environments of LVK sources remain highly uncertain, primarily due to the large distances and limited sky localization accuracy of extragalactic BBH mergers. For example, the typical sky localization accuracy of future third-generation GW detectors is $\sim$$1\,{\rm deg}^2$ (see e.g.,~\citealp{2018PhRvD..98b4029V}), which is still insufficient to localize the host environment at a cosmological distance. Even in mHz GW detections with LISA~\citep{2017arXiv170200786A} (also for Taiji,~\citealp{2020IJMPA..3550075R} and Tianqin,~\citealp{2016CQGra..33c5010L}), the arcminute-level localization accuracy will only allow confident identification of the formation environment for Galactic sources (e.g.,~\citealp{2025arXiv250118682X}).  
Although it is unlikely that a LVK source can be directly associated with a specific host environment via GW detection alone, our analysis demonstrates the theoretical value of incorporating astrophysical priors. By modeling events as previous mergers in the hierarchical triple merger scenario in dense environments such as GCs, we quantify the potential improvement in parameter measurement accuracy if indirect methods, such as population-level spin/mass correlations~\citep[e.g.,][]{2019PhRvL.123r1101Y,2022MNRAS.514.3886M,2023PhRvD.107f3007L,2025ApJ...981..177L} or multi-messenger constraints~\citep[e.g.,][]{2023ApJ...942...99G,2023MNRAS.526.6031V,li2025multimessengerhierarchicaltriplemerger}, can statistically associate mergers with specific formation environment.  
Though host-specific claims remain observationally unverified, our results motivate the development of hybrid frameworks that combine astrophysical environments with GW inference to provide accurate GW PE.

We note that we employ the kick velocity prescription from~\citet{2007ApJ...659L...5C} to estimate remnant black hole recoils, recognizing that numerical relativity fitting formulas may introduce larger uncertainties compared to numerical relativity surrogate models~\citep{2019PhRvR...1c3015V}. For example, the kick velocity distribution of GW170729 shows a bit of discrepancy between our estimates based on numerical relativity fitting formulas and those derived from numerical relativity surrogate models (see Figure~2 in~\citealp{2020PhRvL.124j1104V}).
In particular, current GW observations provide limited information on kick magnitudes due to the lack of tight constraints on the spin parameters of the binary (the Jensen-Shannon divergence is used to quantify the information content on kicks~\citep{61115}, with values for the six GW events considered in this work exceeding the threshold of 0.007 adopted in~\citealp{2021PhRvX..11b1053A}, as shown in Table~2 of~\citealp{2021ApJ...918L..31M}).
However, our primary goal is not to derive precise kick magnitudes, but to demonstrate how environment-dependent escape velocity thresholds can constrain GW parameters under hierarchical merger scenarios (see Equations~\ref{eq:1} and~\ref{eq:2}).  
When interpreting our results, one should consider the simplicity of the kick calculation and the inherent uncertainties in formation environment priors. 
Future applications should incorporate improved kick prescriptions, but our framework highlights the critical role of astrophysical context in interpreting hierarchical triple mergers.

\section{Conclusions}\label{sec:conclusions}
In this study, we present a novel method to constrain the parameters and infer the origin channels, of GW events detected by LVK, leveraging the escape velocities of their host environments. This approach provides more accurate PE for GW events that may be the result of previous mergers in hierarchical triple merger scenarios. Additionally, it supports the search for hierarchical triple mergers using third-generation ground-based GW detectors~\citep{2024ApJ...965...80G}. As examples, we analyze six GW events recently identified as potential precursors in such scenarios~\citep{2021ApJ...907L..48V}.

Applying escape velocity constraints leads to more convergent posterior distributions, which carry richer information to refine our understanding of the physical properties, formation processes, and evolutionary history of the sources. 
Moreover, it enables the exclusion of origin channels that cause noticeable deviations between the LVK posterior distributions and those obtained under escape velocity constraints.

Our results highlight that the most credible candidates for hierarchical triple mergers tend to have lower kick velocities. In particular, by applying escape velocity constraints, the posterior spin distributions are concentrated around zero, aligning with the expected birth spin function of first-generation BHs. Among the analyzed environments, AGN disks and NSCs emerge as the most promising sites for such events (or scenarios). For GCs, we find that if GW170818 originates from GCs, the uncertainty in its posterior primary mass distribution is improved from $\sim$30.1\% to $\sim$22.5\%. On the other hand, GW190512 is unlikely to originate from a GC, based on the significant deviation observed in its constrained posterior distribution. The events also excluded from a GC origin are GW170729 and GW190708 (see Appendix~\ref{appB}).

Previously, we highlighted the challenges of effectively inferring detailed information about the host environment solely from the distribution of BBH merger parameters, especially when multiple formation channels are considered~\citep{2025ApJ...981..177L}. Fortunately, this study demonstrates that the escape velocity of the host environment can be used inversely to probe the nature of GW events, offering a novel approach to address these challenges. Future work can refine this method by incorporating more precise modeling of kick velocity distributions across different environments and their influence on PE. With the deployment of third-generation GW detectors~\citep{2010CQGra..27s4002P,2010CQGra..27h4007P,2017CQGra..34d4001A}, this approach can be extended to a broader range of events, systematically enhancing the search for hierarchical triple mergers. Furthermore, combining this method with multi-messenger astronomy using precise source localization and joint electromagnetic observations offers an opportunity to deepen our understanding of the formation mechanisms, environmental conditions, and cosmological implications of such events~\citep{2025A&A...693A..84K}. Finally, applying this technique to special GW events, such as those with higher masses or complex spin properties, will provide new perspectives for testing general relativity and exploring the evolution of the cosmos.

\section{Acknowledgments}
We would like to thank the referees for their valuable comments and inputs which significantly improved the original manuscript. 
This work is supported by 
the National Key R$\&$D Program of China (2020YFC2201400),
and the National Natural Science Foundation of China (grant No.~11922303). 
This research has made use of data or software obtained from the Gravitational Wave Open Science Center (\url{https://gwosc.org}), a service of the LIGO Scientific Collaboration, the Virgo Collaboration, and KAGRA. This material is based upon work supported by NSF's LIGO Laboratory which is a major facility fully funded by the National Science Foundation, as well as the Science and Technology Facilities Council (STFC) of the United Kingdom, the Max-Planck-Society (MPS), and the State of Niedersachsen/Germany for support of the construction of Advanced LIGO and construction and operation of the GEO600 detector. Additional support for Advanced LIGO was provided by the Australian Research Council. Virgo is funded, through the European Gravitational Observatory (EGO), by the French Centre National de Recherche Scientifique (CNRS), the Italian Istituto Nazionale di Fisica Nucleare (INFN) and the Dutch Nikhef, with contributions by institutions from Belgium, Germany, Greece, Hungary, Ireland, Japan, Monaco, Poland, Portugal, Spain. KAGRA is supported by Ministry of Education, Culture, Sports, Science and Technology (MEXT), Japan Society for the Promotion of Science (JSPS) in Japan; National Research Foundation (NRF) and Ministry of Science and ICT (MSIT) in Korea; Academia Sinica (AS) and National Science and Technology Council (NSTC) in Taiwan.
This analysis was made possible following software packages:
NumPy~\citep{harris2020array}, 
SciPy~\citep{2020SciPy-NMeth}, 
Matplotlib~\citep{2007CSE.....9...90H}, 
IPython~\citep{2007CSE.....9c..21P},
seaborn~\citep{Waskom2021},
and Astropy~\citep{2022ApJ...935..167A}.

\appendix

\section{Posterior parameters}
The posterior parameters in Table~\ref{tab2} are the primary mass $m_1$, the second mass $m_2$, the primary spin magnitude $\chi_1$, the second spin magnitude $\chi_2$, the mass ratio $q$, the chirp mass $\mathcal{M}$, the effective spin $\chi_{\rm eff}$, and the effective precession parameter $\chi_{\rm p}$ of a merger. In particular, 
\begin{equation}
    q = \frac{m_2}{m_1} \,,
\end{equation}
where $m_2 \leq m_1$; 
\begin{equation}
    \mathcal{M} = \frac{(m_1m_2)^{3/5}}{(m_1+m_2)^{1/5}} \,;
\end{equation}
\begin{equation}
    \chi_{\rm eff} = \frac{m_1 \chi_1 {\rm cos}\theta_1+m_2 \chi_2 {\rm cos}\theta_2}{m_1+m_2} \,,
\end{equation}
where $\theta_i$ is the misalignment angle of each BH in a merger;
\begin{equation}
    \chi_{\rm p}={\rm max}\left[\chi_1 {\rm sin}\theta_1,\chi_2 {\rm sin}\theta_2\frac{q(4q+3)}{4+3q}\right].
\end{equation}

\renewcommand{\thefigure}{{S1}}
\begin{figure}
\centering
\includegraphics[width=8cm]{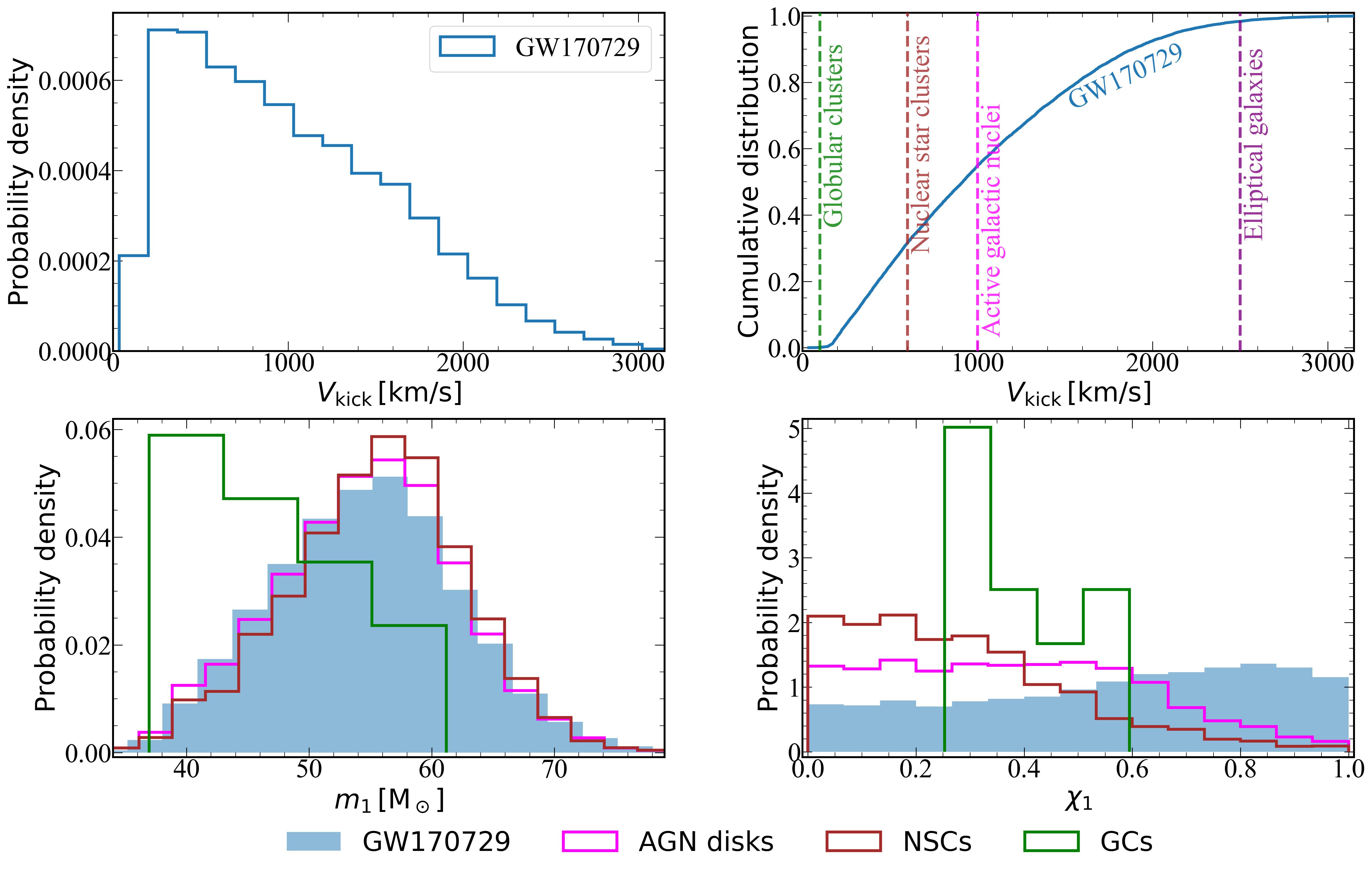}
\caption{
Same as Figure,\ref{kick} and Figure,\ref{GW190512}, but for GW170729, showing the posterior distributions of the kick velocity (top left), primary mass (bottom left), and primary spin (bottom right), with the cumulative distribution shown in the top right. The vertical dotted lines represent the upper bounds of the escape velocity ranges for various types of host environments, provided for comparison.
}
\label{GW170729} 
\end{figure}

\renewcommand{\thefigure}{{S2}}
\begin{figure}
\centering
\includegraphics[width=8cm]{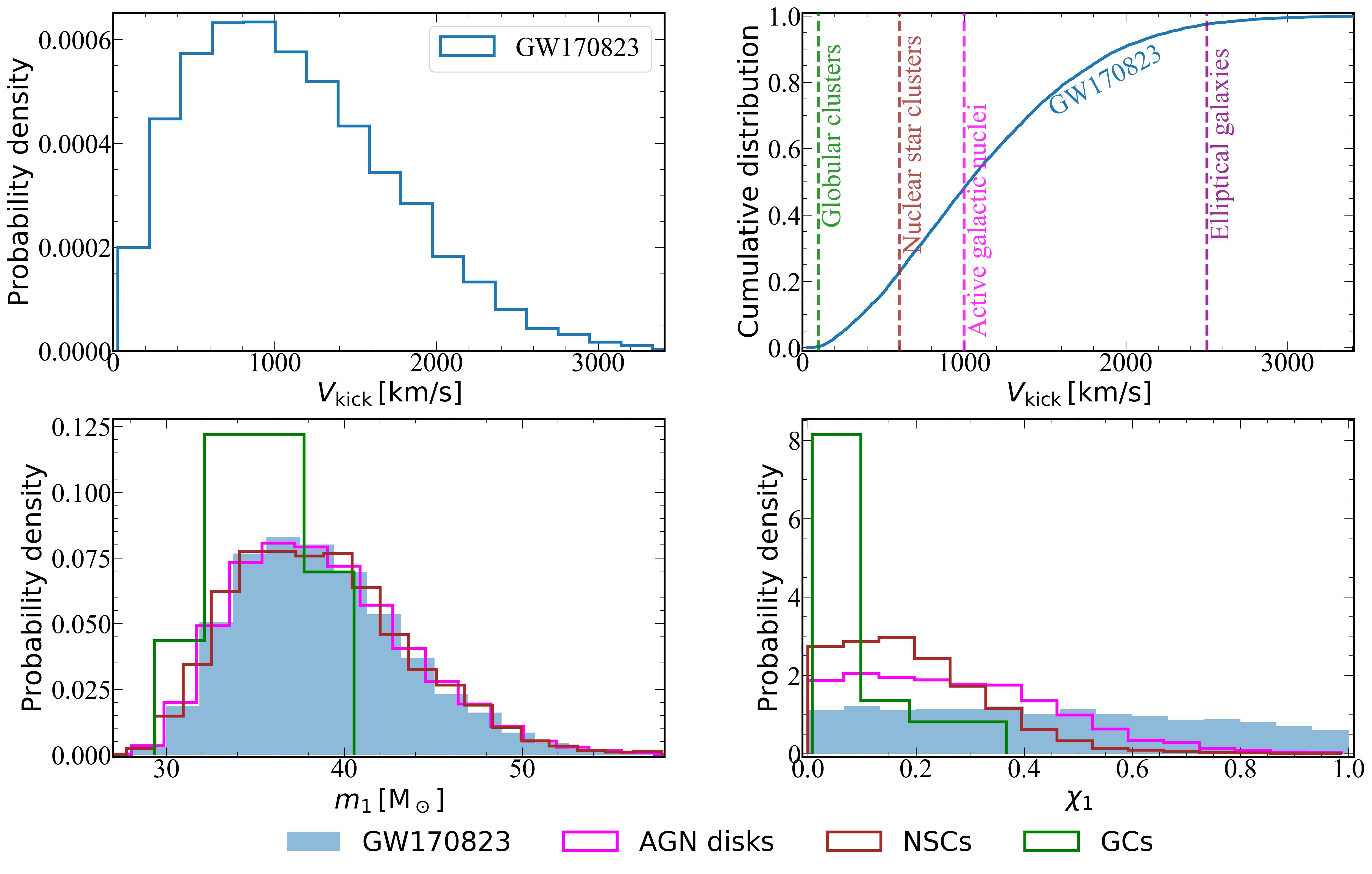}
\caption{
Same as Figure~\ref{GW170729} , but for GW170823.
}
\label{GW170823} 
\end{figure}

\renewcommand{\thefigure}{{S3}}
\begin{figure}
\centering
\includegraphics[width=8cm]{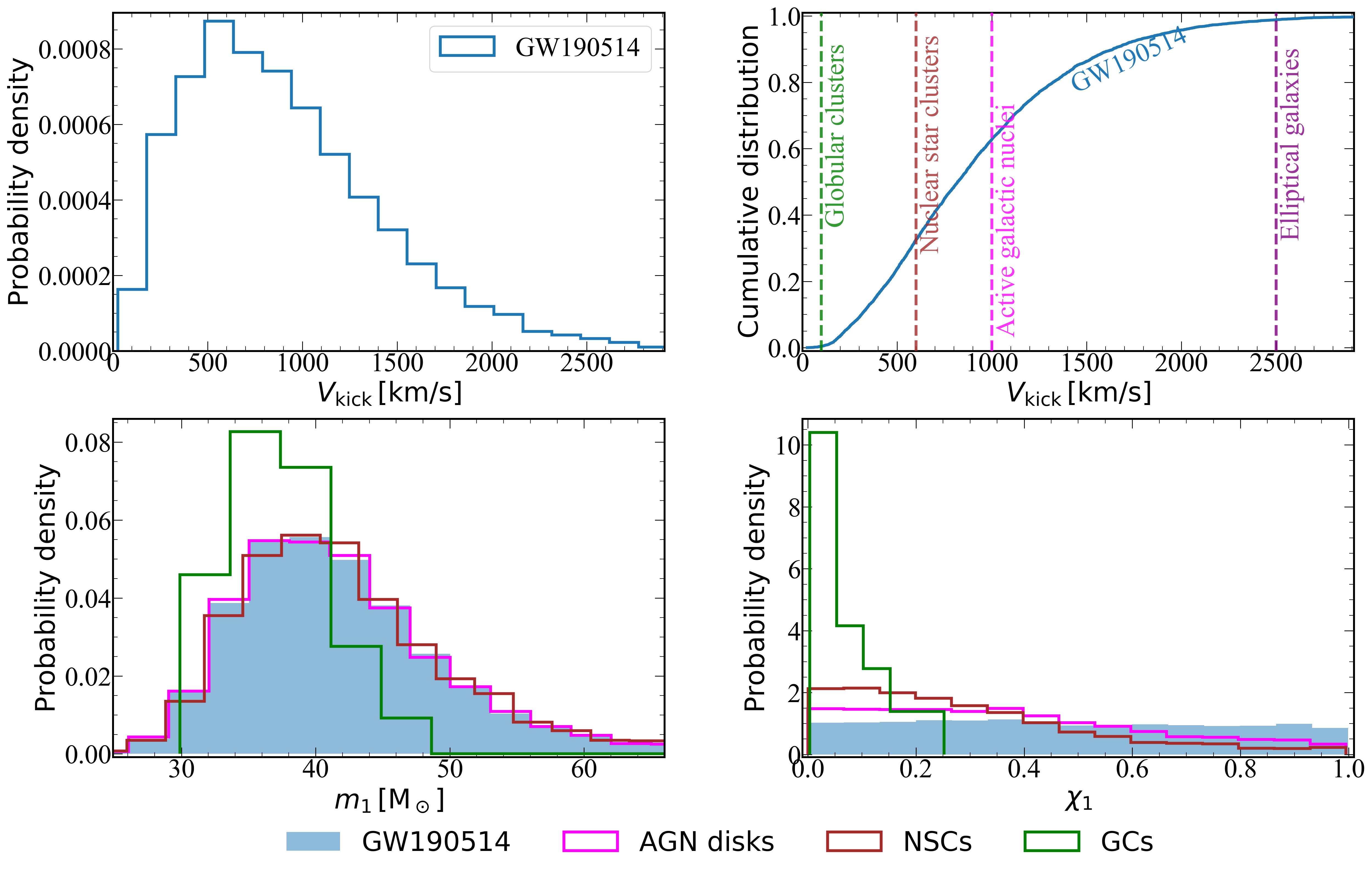}
\caption{
Same as Figure~\ref{GW170823} , but for GW190514.
}
\label{GW190514} 
\end{figure}

\renewcommand{\thefigure}{{S4}}
\begin{figure}
\centering
\includegraphics[width=8cm]{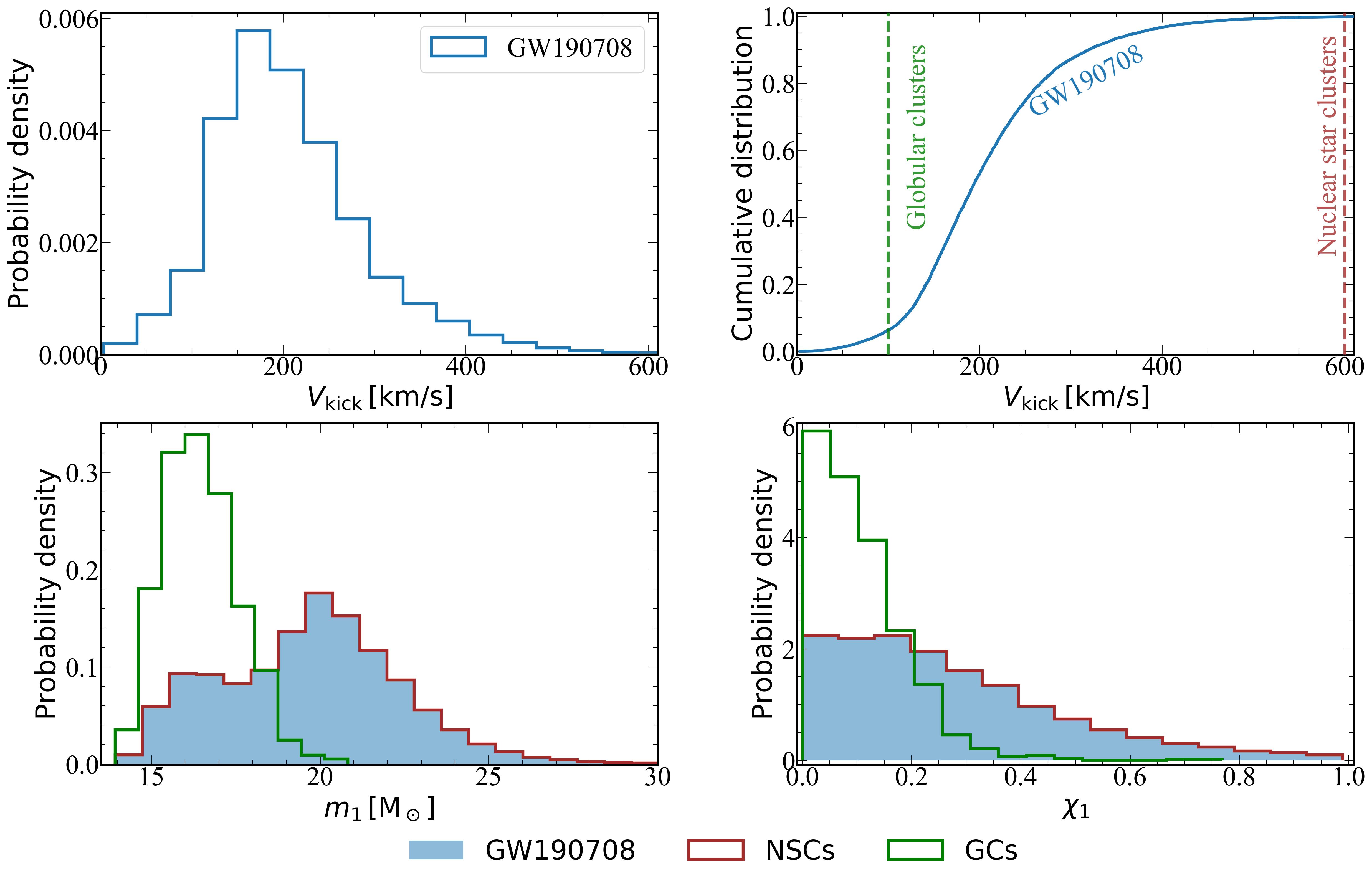}
\caption{
Same as Figure~\ref{GW190514} , but for GW190708.
}
\label{GW190708} 
\end{figure}

\section{Posterior distributions}\label{appB}
Table~\ref{tab3} extends the results presented in Table~\ref{tab1} and Table~\ref{tab2} for the six GW events considered as potential previous mergers in hierarchical triple merger scenarios. The results align well with the analysis discussed in the main text.  

Additionally, we provide posterior distribution plots for GW170729 (Figure~\ref{GW170729}), GW170823 (Figure~\ref{GW170823}), GW190514 (Figure~\ref{GW190514}), and GW190708 (Figure~\ref{GW190708}) that are not displayed in the main text. Notably, the data indicate that GW170729 and GW190708 are unlikely to originate from GCs in the hierarchical triple merger scenario. This conclusion is drawn from their primary mass distributions (see Figures~\ref{GW170729}\&\ref{GW190708}), which show significant deviations from the GWTC-2.1 distributions. Such deviations suggest that the environmental constraints imposed by GCs are inconsistent with the physical characteristics of these events.  

\renewcommand{\thetable}{{S1}}
\begin{table*}
\caption{
Same as Table~\ref{tab2}, but for the six GW events identified as previous mergers in the hierarchical triple merger scenario, obtained under the assumption that the merger takes place inside host environments with the various kick velocities.
}\resizebox{\linewidth}{!}{
\begin{tabular}{cccccccccccc}
\toprule
Event & $V_{\rm esc}\,[{\rm km\,s^{-1}]}$ & $f_{\rm ret}$ & $m_1\,[M_{\odot}]$ & $m_2\,[M_{\odot}]$ & $\chi_1$ & $\chi_2$ & $q$ & $\mathcal{M}\,[M_{\odot}]$ & $\chi_{\rm eff}$ & $\chi_{\rm p}$ \\
\midrule
\multirow{7}{*}{GW170729} & $\cdots$ & $\cdots$ & $54.68^{+12.73}_{-12.77}$ & $30.20^{+11.95}_{-10.16}$ & $0.60^{+0.35}_{-0.53}$ & $0.48^{+0.46}_{-0.44}$ & $0.55^{+0.37}_{-0.22}$ & $34.55^{+6.97}_{-5.72}$ & $0.29^{+0.25}_{-0.33}$ & $0.39^{+0.40}_{-0.29}$ \\
 & $1000$ & $0.5487$ & $55.04^{+11.59}_{-13.24}$ & $27.89^{+13.70}_{-9.12}$ & $0.38^{+0.43}_{-0.34}$ & $0.44^{+0.49}_{-0.40}$ & $0.51^{+0.41}_{-0.20}$ & $33.36^{+7.38}_{-5.31}$ & $0.20^{+0.31}_{-0.28}$ & $0.26^{+0.29}_{-0.19}$ \\
 & $600$ & $0.3164$ & $55.84^{+11.00}_{-13.04}$ & $25.99^{+14.32}_{-7.98}$ & $0.25^{+0.44}_{-0.23}$ & $0.41^{+0.50}_{-0.37}$ & $0.47^{+0.42}_{-0.18}$ & $32.38^{+7.36}_{-4.81}$ & $0.13^{+0.35}_{-0.24}$ & $0.19^{+0.22}_{-0.14}$ \\
 & $300$ & $0.1035$ & $57.10^{+10.86}_{-13.08}$ & $23.63^{+14.58}_{-6.59}$ & $0.14^{+0.42}_{-0.13}$ & $0.28^{+0.54}_{-0.26}$ & $0.42^{+0.43}_{-0.15}$ & $31.28^{+7.15}_{-4.35}$ & $0.05^{+0.40}_{-0.19}$ & $0.11^{+0.15}_{-0.07}$ \\
 & $200$ & $0.0334$ & $57.07^{+11.53}_{-14.61}$ & $23.73^{+12.97}_{-6.94}$ & $0.09^{+0.43}_{-0.08}$ & $0.18^{+0.48}_{-0.17}$ & $0.42^{+0.42}_{-0.17}$ & $31.04^{+7.76}_{-4.10}$ & $0.03^{+0.41}_{-0.15}$ & $0.06^{+0.11}_{-0.05}$ \\
 & $100$ & $0.0009$ & $45.89^{+12.03}_{-8.37}$ & $34.22^{+4.84}_{-4.20}$ & $0.37^{+0.18}_{-0.11}$ & $0.39^{+0.13}_{-0.33}$ & $0.75^{+0.16}_{-0.16}$ & $33.35^{+7.01}_{-2.89}$ & $0.32^{+0.15}_{-0.06}$ & $0.08^{+0.06}_{-0.07}$ \\
 & $50$ & $0.0001$ & $\cdots$ & $\cdots$ & $\cdots$ & $\cdots$ & $\cdots$ & $\cdots$ & $\cdots$ & $\cdots$ \\
\multirow{7}{*}{GW170818} & $\cdots$ & $\cdots$ & $34.78^{+6.52}_{-4.19}$ & $27.60^{+4.15}_{-5.10}$ & $0.52^{+0.41}_{-0.47}$ & $0.48^{+0.46}_{-0.42}$ & $0.80^{+0.18}_{-0.24}$ & $26.83^{+2.32}_{-2.00}$ & $-0.06^{+0.19}_{-0.22}$ & $0.56^{+0.34}_{-0.41}$ \\
 & $1000$ & $1.0000$ & $34.78^{+6.52}_{-4.19}$ & $27.60^{+4.15}_{-5.10}$ & $0.52^{+0.41}_{-0.47}$ & $0.48^{+0.46}_{-0.42}$ & $0.80^{+0.18}_{-0.24}$ & $26.83^{+2.32}_{-2.00}$ & $-0.06^{+0.19}_{-0.22}$ & $0.56^{+0.34}_{-0.41}$ \\
 & $600$ & $1.0000$ & $34.78^{+6.52}_{-4.19}$ & $27.60^{+4.15}_{-5.10}$ & $0.52^{+0.41}_{-0.47}$ & $0.48^{+0.46}_{-0.42}$ & $0.80^{+0.18}_{-0.24}$ & $26.83^{+2.32}_{-2.00}$ & $-0.06^{+0.19}_{-0.22}$ & $0.56^{+0.34}_{-0.41}$ \\
 & $300$ & $0.8386$ & $34.65^{+6.34}_{-4.12}$ & $27.58^{+4.13}_{-5.06}$ & $0.45^{+0.44}_{-0.40}$ & $0.44^{+0.49}_{-0.39}$ & $0.80^{+0.17}_{-0.24}$ & $26.77^{+2.30}_{-1.96}$ & $-0.06^{+0.18}_{-0.22}$ & $0.51^{+0.35}_{-0.38}$ \\
 & $200$ & $0.4483$ & $34.34^{+6.02}_{-4.00}$ & $27.71^{+4.03}_{-4.96}$ & $0.30^{+0.42}_{-0.27}$ & $0.39^{+0.48}_{-0.35}$ & $0.81^{+0.17}_{-0.23}$ & $26.70^{+2.28}_{-1.90}$ & $-0.06^{+0.17}_{-0.21}$ & $0.37^{+0.35}_{-0.28}$ \\
 & $100$ & $0.0882$ & $33.31^{+4.14}_{-3.36}$ & $28.59^{+3.47}_{-3.93}$ & $0.17^{+0.35}_{-0.16}$ & $0.24^{+0.43}_{-0.21}$ & $0.87^{+0.12}_{-0.18}$ & $26.72^{+2.44}_{-1.87}$ & $-0.04^{+0.14}_{-0.21}$ & $0.19^{+0.35}_{-0.14}$ \\
 & $50$ & $0.0108$ & $32.01^{+2.81}_{-2.33}$ & $29.64^{+2.67}_{-2.43}$ & $0.10^{+0.30}_{-0.09}$ & $0.13^{+0.29}_{-0.11}$ & $0.94^{+0.05}_{-0.11}$ & $26.82^{+2.10}_{-1.65}$ & $-0.01^{+0.10}_{-0.16}$ & $0.11^{+0.26}_{-0.08}$ \\
\multirow{7}{*}{GW170823} & $\cdots$ & $\cdots$ & $38.26^{+9.52}_{-6.15}$ & $28.98^{+6.55}_{-7.78}$ & $0.44^{+0.48}_{-0.39}$ & $0.43^{+0.49}_{-0.39}$ & $0.78^{+0.20}_{-0.30}$ & $28.58^{+4.55}_{-3.33}$ & $0.05^{+0.21}_{-0.22}$ & $0.47^{+0.41}_{-0.35}$ \\
 & $1000$ & $0.4814$ & $38.23^{+9.43}_{-6.32}$ & $28.55^{+6.72}_{-8.11}$ & $0.26^{+0.37}_{-0.23}$ & $0.33^{+0.47}_{-0.30}$ & $0.76^{+0.21}_{-0.30}$ & $28.24^{+4.64}_{-3.34}$ & $0.03^{+0.20}_{-0.22}$ & $0.29^{+0.27}_{-0.21}$ \\
 & $600$ & $0.2274$ & $38.20^{+9.48}_{-6.49}$ & $28.51^{+6.35}_{-7.78}$ & $0.18^{+0.28}_{-0.16}$ & $0.24^{+0.40}_{-0.21}$ & $0.76^{+0.21}_{-0.30}$ & $28.20^{+4.51}_{-3.16}$ & $0.02^{+0.19}_{-0.19}$ & $0.19^{+0.21}_{-0.13}$ \\
 & $300$ & $0.0694$ & $37.81^{+9.37}_{-6.14}$ & $28.58^{+6.12}_{-7.74}$ & $0.10^{+0.24}_{-0.09}$ & $0.14^{+0.26}_{-0.12}$ & $0.77^{+0.20}_{-0.30}$ & $28.10^{+4.23}_{-3.10}$ & $0.01^{+0.16}_{-0.16}$ & $0.10^{+0.15}_{-0.07}$ \\
 & $200$ & $0.0294$ & $37.39^{+9.49}_{-5.84}$ & $28.99^{+5.62}_{-7.82}$ & $0.07^{+0.23}_{-0.06}$ & $0.10^{+0.26}_{-0.09}$ & $0.79^{+0.19}_{-0.31}$ & $28.14^{+4.05}_{-2.97}$ & $0.01^{+0.18}_{-0.12}$ & $0.07^{+0.13}_{-0.04}$ \\
 & $100$ & $0.0033$ & $35.63^{+4.37}_{-5.14}$ & $30.61^{+4.10}_{-3.12}$ & $0.04^{+0.29}_{-0.03}$ & $0.04^{+0.27}_{-0.04}$ & $0.88^{+0.11}_{-0.14}$ & $28.59^{+3.29}_{-2.95}$ & $0.01^{+0.18}_{-0.10}$ & $0.04^{+0.09}_{-0.03}$ \\
 & $50$ & $0.0003$ & $\cdots$ & $\cdots$ & $\cdots$ & $\cdots$ & $\cdots$ & $\cdots$ & $\cdots$ & $\cdots$ \\
\multirow{7}{*}{GW190512} & $\cdots$ & $\cdots$ & $23.15^{+5.64}_{-5.61}$ & $12.53^{+3.49}_{-2.55}$ & $0.20^{+0.49}_{-0.18}$ & $0.40^{+0.51}_{-0.37}$ & $0.54^{+0.36}_{-0.18}$ & $14.56^{+1.36}_{-0.94}$ & $0.02^{+0.13}_{-0.14}$ & $0.26^{+0.41}_{-0.20}$ \\
 & $1000$ & $0.8526$ & $23.50^{+5.43}_{-5.66}$ & $12.38^{+3.45}_{-2.47}$ & $0.17^{+0.34}_{-0.15}$ & $0.37^{+0.51}_{-0.33}$ & $0.53^{+0.35}_{-0.17}$ & $14.57^{+1.34}_{-0.95}$ & $0.03^{+0.12}_{-0.13}$ & $0.22^{+0.27}_{-0.17}$ \\
 & $600$ & $0.5771$ & $23.84^{+5.24}_{-5.68}$ & $12.23^{+3.51}_{-2.40}$ & $0.14^{+0.22}_{-0.12}$ & $0.29^{+0.49}_{-0.26}$ & $0.51^{+0.34}_{-0.16}$ & $14.59^{+1.33}_{-0.97}$ & $0.03^{+0.12}_{-0.12}$ & $0.17^{+0.17}_{-0.13}$ \\
 & $300$ & $0.2219$ & $23.60^{+5.21}_{-5.67}$ & $12.40^{+3.62}_{-2.56}$ & $0.09^{+0.15}_{-0.08}$ & $0.16^{+0.35}_{-0.14}$ & $0.52^{+0.35}_{-0.17}$ & $14.62^{+1.33}_{-0.98}$ & $0.02^{+0.12}_{-0.09}$ & $0.09^{+0.11}_{-0.07}$ \\
 & $200$ & $0.0844$ & $22.63^{+5.30}_{-4.90}$ & $12.93^{+3.35}_{-2.82}$ & $0.06^{+0.13}_{-0.05}$ & $0.09^{+0.23}_{-0.08}$ & $0.57^{+0.33}_{-0.19}$ & $14.67^{+1.29}_{-1.02}$ & $0.01^{+0.11}_{-0.08}$ & $0.05^{+0.08}_{-0.04}$ \\
 & $100$ & $0.0046$ & $18.84^{+2.52}_{-2.01}$ & $15.68^{+2.26}_{-1.61}$ & $0.03^{+0.10}_{-0.03}$ & $0.04^{+0.08}_{-0.03}$ & $0.84^{+0.13}_{-0.12}$ & $15.08^{+1.09}_{-1.09}$ & $-0.00^{+0.07}_{-0.05}$ & $0.03^{+0.08}_{-0.02}$ \\
 & $50$ & $0.0002$ & $\cdots$ & $\cdots$ & $\cdots$ & $\cdots$ & $\cdots$ & $\cdots$ & $\cdots$ & $\cdots$ \\
\multirow{7}{*}{GW190514} & $\cdots$ & $\cdots$ & $40.87^{+17.31}_{-9.30}$ & $28.35^{+9.96}_{-10.05}$ & $0.47^{+0.47}_{-0.42}$ & $0.48^{+0.47}_{-0.43}$ & $0.71^{+0.26}_{-0.35}$ & $29.14^{+8.08}_{-5.42}$ & $-0.08^{+0.29}_{-0.35}$ & $0.45^{+0.42}_{-0.33}$ \\
 & $1000$ & $0.6290$ & $40.86^{+17.67}_{-9.43}$ & $27.64^{+9.92}_{-10.21}$ & $0.35^{+0.52}_{-0.31}$ & $0.40^{+0.51}_{-0.37}$ & $0.69^{+0.27}_{-0.36}$ & $28.68^{+7.74}_{-5.42}$ & $-0.10^{+0.27}_{-0.37}$ & $0.33^{+0.35}_{-0.24}$ \\
 & $600$ & $0.3263$ & $40.99^{+19.10}_{-9.30}$ & $27.17^{+10.05}_{-10.79}$ & $0.24^{+0.53}_{-0.22}$ & $0.32^{+0.55}_{-0.29}$ & $0.68^{+0.29}_{-0.38}$ & $28.42^{+7.86}_{-5.57}$ & $-0.09^{+0.25}_{-0.39}$ & $0.22^{+0.27}_{-0.16}$ \\
 & $300$ & $0.0914$ & $40.91^{+20.91}_{-9.67}$ & $27.27^{+10.36}_{-12.10}$ & $0.13^{+0.37}_{-0.12}$ & $0.19^{+0.57}_{-0.17}$ & $0.70^{+0.27}_{-0.44}$ & $28.40^{+8.01}_{-5.96}$ & $-0.05^{+0.17}_{-0.37}$ & $0.11^{+0.24}_{-0.08}$ \\
 & $200$ & $0.0355$ & $39.77^{+19.90}_{-8.89}$ & $28.86^{+8.60}_{-12.32}$ & $0.09^{+0.28}_{-0.08}$ & $0.12^{+0.63}_{-0.11}$ & $0.75^{+0.22}_{-0.47}$ & $28.88^{+6.98}_{-6.00}$ & $-0.02^{+0.14}_{-0.35}$ & $0.08^{+0.16}_{-0.06}$ \\
 & $100$ & $0.0042$ & $37.63^{+38.10}_{-5.74}$ & $31.27^{+9.92}_{-23.44}$ & $0.05^{+0.13}_{-0.05}$ & $0.08^{+0.68}_{-0.07}$ & $0.86^{+0.12}_{-0.76}$ & $30.25^{+6.11}_{-10.76}$ & $-0.02^{+0.11}_{-0.16}$ & $0.04^{+0.07}_{-0.03}$ \\
 & $50$ & $0.0004$ & $\cdots$ & $\cdots$ & $\cdots$ & $\cdots$ & $\cdots$ & $\cdots$ & $\cdots$ & $\cdots$ \\
\multirow{7}{*}{GW190708} & $\cdots$ & $\cdots$ & $19.81^{+4.26}_{-4.34}$ & $11.61^{+3.08}_{-1.96}$ & $0.23^{+0.47}_{-0.20}$ & $0.34^{+0.53}_{-0.30}$ & $0.58^{+0.36}_{-0.18}$ & $13.05^{+0.86}_{-0.59}$ & $0.05^{+0.10}_{-0.10}$ & $0.26^{+0.44}_{-0.20}$ \\
 & $1000$ & $1.0000$ & $19.81^{+4.26}_{-4.34}$ & $11.61^{+3.08}_{-1.96}$ & $0.23^{+0.47}_{-0.20}$ & $0.34^{+0.53}_{-0.30}$ & $0.58^{+0.36}_{-0.18}$ & $13.05^{+0.86}_{-0.59}$ & $0.05^{+0.10}_{-0.10}$ & $0.26^{+0.44}_{-0.20}$ \\
 & $600$ & $0.9983$ & $19.81^{+4.26}_{-4.34}$ & $11.61^{+3.09}_{-1.96}$ & $0.23^{+0.47}_{-0.20}$ & $0.33^{+0.52}_{-0.30}$ & $0.58^{+0.36}_{-0.18}$ & $13.05^{+0.86}_{-0.59}$ & $0.05^{+0.11}_{-0.10}$ & $0.26^{+0.44}_{-0.20}$ \\
 & $300$ & $0.8709$ & $19.91^{+4.29}_{-4.37}$ & $11.55^{+3.10}_{-1.94}$ & $0.20^{+0.32}_{-0.18}$ & $0.30^{+0.50}_{-0.27}$ & $0.58^{+0.36}_{-0.18}$ & $13.06^{+0.86}_{-0.60}$ & $0.06^{+0.10}_{-0.10}$ & $0.23^{+0.30}_{-0.18}$ \\
 & $200$ & $0.5285$ & $19.56^{+4.36}_{-4.11}$ & $11.76^{+3.03}_{-2.06}$ & $0.17^{+0.22}_{-0.15}$ & $0.22^{+0.34}_{-0.20}$ & $0.60^{+0.35}_{-0.19}$ & $13.08^{+0.87}_{-0.62}$ & $0.05^{+0.10}_{-0.09}$ & $0.17^{+0.21}_{-0.13}$ \\
 & $100$ & $0.0630$ & $16.37^{+2.09}_{-1.47}$ & $14.07^{+1.40}_{-1.56}$ & $0.09^{+0.17}_{-0.08}$ & $0.10^{+0.19}_{-0.09}$ & $0.86^{+0.12}_{-0.16}$ & $13.16^{+0.86}_{-0.67}$ & $0.00^{+0.06}_{-0.06}$ & $0.10^{+0.16}_{-0.08}$ \\
 & $50$ & $0.0125$ & $15.70^{+1.46}_{-1.00}$ & $14.62^{+1.07}_{-0.86}$ & $0.05^{+0.09}_{-0.04}$ & $0.06^{+0.09}_{-0.05}$ & $0.94^{+0.06}_{-0.10}$ & $13.20^{+0.88}_{-0.66}$ & $-0.00^{+0.04}_{-0.05}$ & $0.06^{+0.07}_{-0.05}$ \\
\bottomrule
\end{tabular}}
\label{tab3}
\begin{tablenotes} 
\item 
{\bf Note.}
Column 1: Name of the GW events.  
Column 2: Escape velocities of the host environments.  
Column 3: Retention fractions of the posterior samples that have been retained, assuming the merger takes place inside the host environment given the escape velocities.  
Columns 4-11: Median and 90\% credible intervals for the corresponding posterior distributions of $m_1$, $m_2$, $\chi_1$, $\chi_2$, $q$, $\mathcal{M}$, $\chi_{\rm eff}$, and $\chi_{\rm p}$.  
`$\cdots$' in both $V_{\rm kick}$ and $f_{\rm ret}$ indicates that the posterior distributions are from GWTC-2.1, while other cases indicate that the number of posterior samples is too small to provide reliable median and credible intervals for the distribution.  
\end{tablenotes}
\end{table*}

\clearpage

\end{document}